\documentclass[aps,prl,twocolumn,showpacs,preprintnumbers,amsmath,amssymb]{revtex4-1}
\usepackage{latexsym}
\usepackage{graphicx}
\usepackage{graphics}
\usepackage[T1]{fontenc}
\usepackage{multirow}
\usepackage[british,UKenglish,USenglish,english,american]{babel}
\usepackage{color}
\usepackage{booktabs}

\newcommand{\ket}[1]{\mbox{$ | #1 \rangle $}}

\begin{document}

\author{Panagiotis Vergyris$^1$, Florent Mazeas$^{1,\dagger}$, Elie Gouzien$^1$, Laurent Labont\'e$^1$, Olivier Alibart$^1$, S\'ebastien Tanzilli$^1$}
\author{Florian Kaiser$^{1,\ddagger}$}
\email{f.kaiser@physik.uni-stuttgart.de}

\affiliation{$^1$Universit\'e C\^ote d'Azur, Institut de Physique de Nice (INPHYNI), CNRS UMR 7010, Parc Valrose, 06108 Nice Cedex 2, (France)\\
$^{\dagger}$Now at Ecole Polytechnique F\'ed\'erale de Lausanne, Photonic Systems Laboratory (PHOSL), STI-IEL, Station 11, Lausanne CH-1015, (Switzerland)\\
$^{\ddagger}$Now at 3rd Physics Institute of Physics, University of Stuttgart and Center for Integrated Quantum Science and Technology, IQST, Stuttgart, (Germany)}

\title{Fibre based hyperentanglement generation for dense wavelength division multiplexing}

\begin{abstract}
Entanglement is a key resource in quantum information science and associated emerging technologies. Photonic systems offer a large range of exploitable entanglement degrees of freedom such as frequency, time, polarization, and spatial modes. Hyperentangled photons exploit multiple degrees of freedom simultaneously to enhance the performance of quantum information protocols.
Here, we report a fully guided-wave approach for generating polarization and energy-time hyperentangled photons at telecom wavelengths. Moreover, by demultiplexing the broadband emission spectrum of the source into five standard telecom channel pairs, we demonstrate compliance with fibre network standards and improve the effective bit rate capacity of the quantum channel up to one order of magnitude. In all channel pairs, we observe a violation of a generalised Bell inequality by more than 27 standard deviations, underlining the relevance of our approach.
\end{abstract}

\maketitle

\begin{center}
\textit{Introduction}
\end{center}
Precise engineering and control of entanglement has led to remarkable advances in quantum information science. Photonic entanglement is advantageous over classical means for securing communication~\cite{Scarani_QKD_2009}, solving computational problems faster~\cite{Steane_qcomp_1998,Barz_qu_comp_2012}, and in high-precision optical sensing~\cite{Giovannetti_qmet_2011,Wolfgramm_probing_2012,Kaiser_Dispersion_2017}.
It has been shown that a practical quantum advantage can be reached with systems comprising a few tens of qubits~\cite{Aaronson_qcomp_2011}. 
To engineer and access the resulting Hilbert space, one way is to coherently superpose $n \gg 10$ photons emitted from quantum dots~\cite{Somaschi_single_photons_2016,Wang_boson_sampling_2017} or from nonlinear wavemixing sources~\cite{Wang_10photon_2016,Wang_hyperFWM_2018}. To reduce the associated challenges on the photon generation side~\cite{Stanley_2014_QDs_fluctuations}, less photons can be entangled over more degrees of freedom (DOF), referred to as hyperentanglement~\cite{Kwiat_hyper_1997,Barreiro_hyper_2005,Barbieri_hyper_2006,Krenn_qkd_2016,Kues_hyper_chip_2017}.
Photons are excellent candidates for carrying hyperentanglement due to a large variety of exploitable DOFs.
Hyperentangled states are advantageous over their single-observable counterparts in many ways. They lead to a stronger violation of local realist theories, making them less sensitive against decoherence~\cite{Barbieri_hyper_2006}. From the applied side, complete photonic Bell state analysis can be implemented~\cite{Wei_complete_2007}, having immediate repercussions in teleportation-based quantum networking. Additionally, detection failure on one DOF does not necessarily stop networking activity, as faithful entanglement transport is still achieved over all other DOFs.

Considering two DOFs, hyperentanglement quality and usability can be inferred through a generalized Bell inequality, based on two Bell operators $\beta_{1,2}$ $-$ one for each DOF~\cite{Barbieri_hyper_2006}. For theories admitting local elements of reality, it is $|\langle \beta_{1,2} \rangle | \leq 2$, while quantum physics permits reaching $| \langle \beta_{1,2} \rangle | = 2\,\sqrt{2}$.
Previous experiments inferred $| \langle \beta_{1,2} \rangle |$ through sequential measurements on each individual DOF with adapted analyzers~\cite{Barreiro_hyper_2005,Barreiro_hyper_2005,Suo_hyper_2015}. However, this strategy does not provide evidence for an application-relevant quantum advantage. Crucially, a faithful analysis necessitates controlling (and stabilising) all analyzers simultaneously, and showing that they do not influence each other.
Provided that simultaneous measurements on all observables are possible, one can violate a generalized Bell inequality via the operator $\beta = \beta_1 \otimes \beta_2$~\cite{Barbieri_hyper_2006}. Here, local realistic theories predict $|\langle \beta \rangle | \leq 4$, and quantum physics permits reaching a twice as high value, $| \langle \beta \rangle | = 8$.
It is important to note that the only pure states saturating the generalized Bell inequality with $| \langle \beta \rangle | = 8$ are those that achieve $| \langle \beta_{1,2} \rangle | = 2 \sqrt{2}$ simultaneously.
Additionally, one can estimate the fidelity of the generated hyperentangled state using likelihood maximisation tomography~\cite{Qtomo}.

Here, we use those criteria to benchmark our practical and fully guided-wave source of telecom-wavelength hyperentangled photon pairs.
In our experimental configuration, we infer all hyperentangled observables simultaneously without replacing components on the setup. This allows to faithfully quantify the suitability of our approach for quantum networking applications.
We exploit polarization and energy-time DOFs as they can be efficiently guided in standard fiber optical networks.
To further enhance the quantum channel networking capacity, we demonstrate also compliance with telecommunication standards by analyzing hyperentanglement in five dense wavelength division multiplexed channel pairs~\citep{Aktas_DWDM_2016}. The combination of hyperentanglement and multiplexing permits to increase the channel capacity up to one order of magnitude compared to ordinary (single-observable) entanglement distribution.
This paves the way for more efficient quantum information protocols, notably in quantum communication and computation~\cite{Fu_Guo_2017}. 
\begin{center}
\textit{Methods}
\end{center}
\begin{figure*}
\includegraphics[width=2\columnwidth]{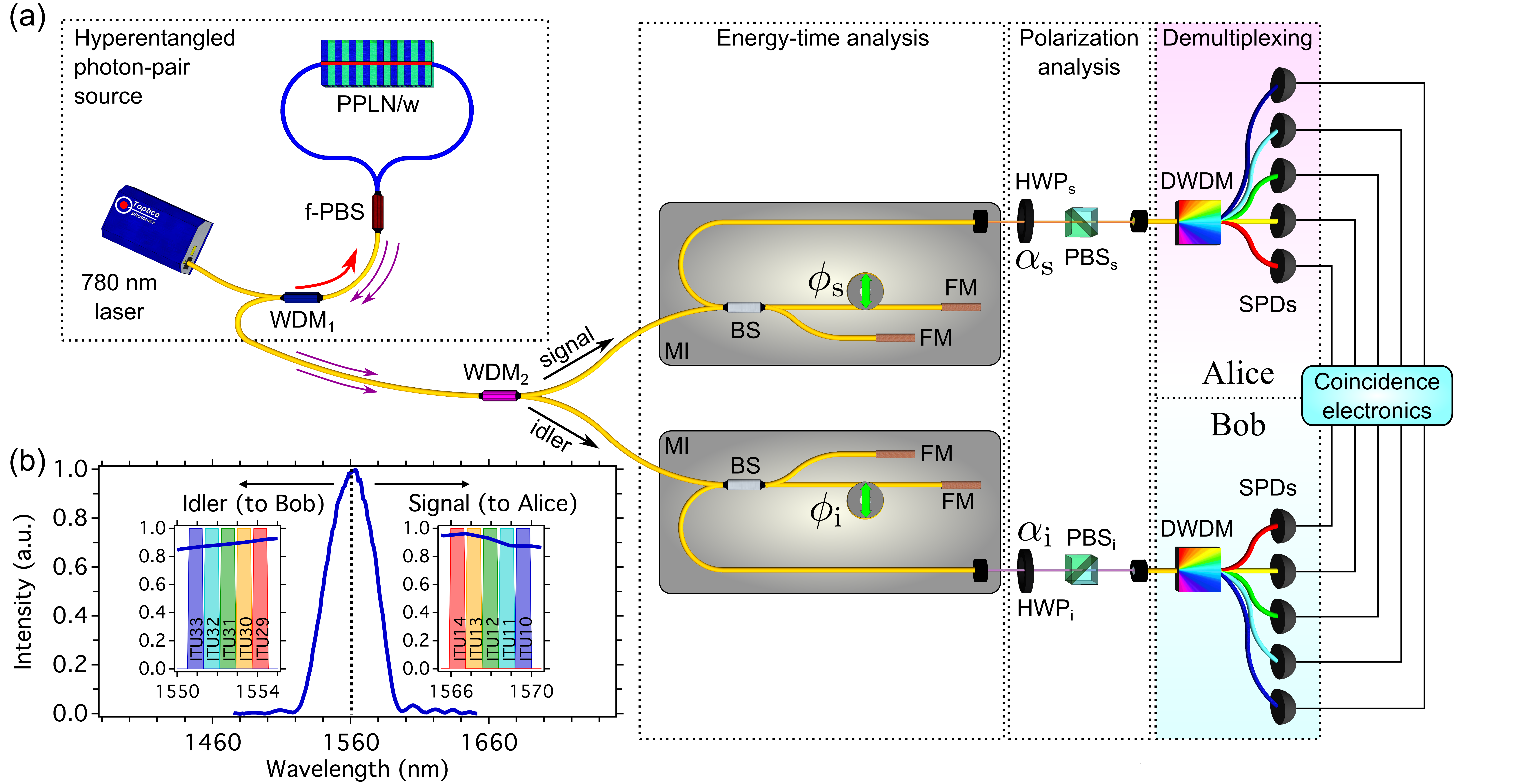}
\caption{\textbf{Setup for the generation and analysis of hyperentanglement.} (a) The photon pair source is based on a fully fibred nonlinear Sagnac interferometer~\cite{Vergyris_sagnac_2017}. Blue lines indicate PMFs. After deterministic separation of signal and idler photons (WDM$_2$), they are sent to their respective hyperentanglement analyzers. The latter are composed of an unbalanced MI for energy-time entanglement analysis and a polarization state analyzer made of a HWP and PBS. Thereafter, the photons are spectrally demultiplexed using DWDMs and detected using standard SPDs. Hyperentanglement in wavelength anticorrelated channel pairs is revealed through coincidence detection. (b) Measured photon pair emission spectrum centred at 1560\,nm. The insets show how Alice's and Bob's photons are spectrally demultiplexed accordingly to the 100\,GHz ITU channel grid.\label{sagnac}}
\end{figure*}
As depicted in \figurename~\ref{sagnac}(a), our scheme is based on a fully guided-wave nonlinear Sagnac interferometer (see also~\cite{Vergyris_sagnac_2017}). Our goal is to generate a hyperentangled two-photon state of the form $\ket{ \Psi}_k = \tfrac{\big( | H\rangle |H\rangle+|V\rangle |V\rangle \big)_k \otimes \big( |E \rangle |E \rangle+ |L \rangle |L \rangle \big)_k}{2}$.
Here, $H$ and $V$ represent horizontal and vertical photon polarization modes, and $E$ and $L$ denote early and late emission times of an energy anticorrelated photon pair~\cite{Franson}.
The index $k$ labels the wavelength channel pair in which the pair is generated.
As a pump, we use a wavelength-stabilized fiber-coupled 780\,nm continuous-wave laser, which is sent through a wavelength division multiplexer (WDM$_1$) to the nonlinear fiber Sagnac loop in which a fiber polarizing beam-splitter (f-PBS) defines the input and output.
After the f-PBS, horizontally (vertically) polarized light propagates in the clockwise (counter-clockwise) direction through polarization maintaining fibers (PMF). One of the PMFs is physically rotated by $90^{\circ}$ such that vertically polarized 780\,nm light pumps a 3.8\,cm long periodically poled lithium niobate waveguide (PPLN/w) from both sides simultaneously. Inside the PPLN/w, pump photons are converted to vertically polarized signal (s) and idler (i) photon pair contributions in both directions through type-0 SPDC.
As shown in \figurename~\ref{sagnac}(b), the photon pair emission spectrum shows a bandwidth of about 40\,nm centered at the degenerated wavelength of 1560\,nm. We now define that signal (idler) photons are above (below) degeneracy.
After the PPLN/w, the pair contributions are coupled back into the PMF, subsequently overlapped at the f-PBS, and separated from the pump at WDM$_1$. By precisely adjusting the polarization of the pump laser, a maximally polarization entangled Bell state is generated: $| \Phi^{+}_{\rm pol} \rangle =\frac{1}{\sqrt{2}} \left( |H \rangle_{\rm s} |H \rangle_{\rm i} + |V \rangle_{\rm s} |V \rangle_{\rm i} \right)$~\cite{Vergyris_sagnac_2017}.
The paired photons are further deterministically separated as a function of their wavelengths using a standard telecom C/L-band splitter (WDM$_2$), and sent to Alice and Bob.
We project the photons' polarization state onto angles $\alpha_{\rm s}$ and $\alpha_{\rm i}$, respectively using a half-wave plate (HWP), a PBS, followed by single-photon detectors (SPD, IDQ220, 20\% efficiency, 100\,ps timing resolution). Coincidence counting allows then to reveal non local correlations.\\
The second DOF of our photon pairs is energy-time entanglement which is mediated through the energy conservation of the SPDC process. As the energy of each generated photon pair must  equal the energy of one pump laser photon, the (vacuum) wavelengths of the involved photons are related by: $\lambda_{\rm p}^{-1} = \lambda_{\rm s}^{-1} + \lambda_{\rm i}^{-1}$. Here, the subscript p stands for the pump photon. Energy-time entanglement is revealed using unbalanced Michelson interferometers (MI) in the Franson configuration~\cite{Franson}. For optimal passive stability, our home-made MIs are made fabricated using fused fiber beam-splitters (BS), and Faraday mirrors (FM) at which photons are reflected back to the BS~\cite{Kaiser_optimal_2016}. Moreover, both MIs are actively stabilized using a 1560\,nm reference laser, and a piezoelectric fibre stretcher in the longer arm~\cite{Kaiser_source2}.
To avoid undesired single-photon interference, each MI has a travel time difference of 300\,ps, which is much larger than the single-photon coherence time, $\tau_{\rm c} = 5\rm\,ps$. However, we adjust both interferometers to have identical travel time differences within $\pm 0.03\rm\,ps$, so as to observe and maximise higher-order interference for simultaneously arriving photons~\cite{Franson,Kaiser_source2,Aktas_DWDM_2016}.
Postselection of simultaneously arriving photons thus results in an energy-time entangled state:
$| \Psi_{\rm e-t} \rangle =\frac{1}{\sqrt{2}} \left( |E \rangle_{\rm s} |E \rangle_{\rm i} + e^{ i \left( \phi_{\rm s} + \phi_{\rm i}\right)} | L \rangle_{\rm s} | L \rangle_{\rm i} \right)$. Here $\phi_{\rm s}$ and $\phi_{\rm i}$ are phase terms depending on the path length difference of the MIs, which can be precisely adjusted to arbitrary settings using the active stabilization system.\\
As our setup provides polarization and energy-time entanglement simultaneously, the resulting overall quantum state reads
\begin{equation}
\ket{ \psi}=\tfrac{\big( |E \rangle_{\rm i} |E\rangle_{\rm s}+e^{ i \left( \phi_{\rm s} + \phi_{\rm i}\right)} |L\rangle_{\rm i} |L\rangle_{\rm s} \big)\otimes \big( | H\rangle_{\rm i} |H\rangle_{\rm s}+|V\rangle_{\rm i} |V\rangle_{\rm s} \big)}{2},\label{eq_1}
\end{equation}
which covers a 16-dimensional Hilbert space.

To even further increase the quantum channel capacity, we exploit standard telecom dense wavelength division multiplexing (DWDM).
As shown in \figurename~\ref{sagnac}(b), photon pairs are created pairwise symmetrically (anti-correlated) around the degenerate wavelength of 1560\,nm.
We exploit this to demultiplex the spectrum into the channel pairs ITU10$-$33, ITU11$-$32, ITU12$-$31, ITU13$-$30, and ITU14$-$29, according to the International Telecommunication Union (ITU) standards in the 100\,GHz grid~\cite{ITU}.

\begin{figure}[t]
\includegraphics[width=1\columnwidth]{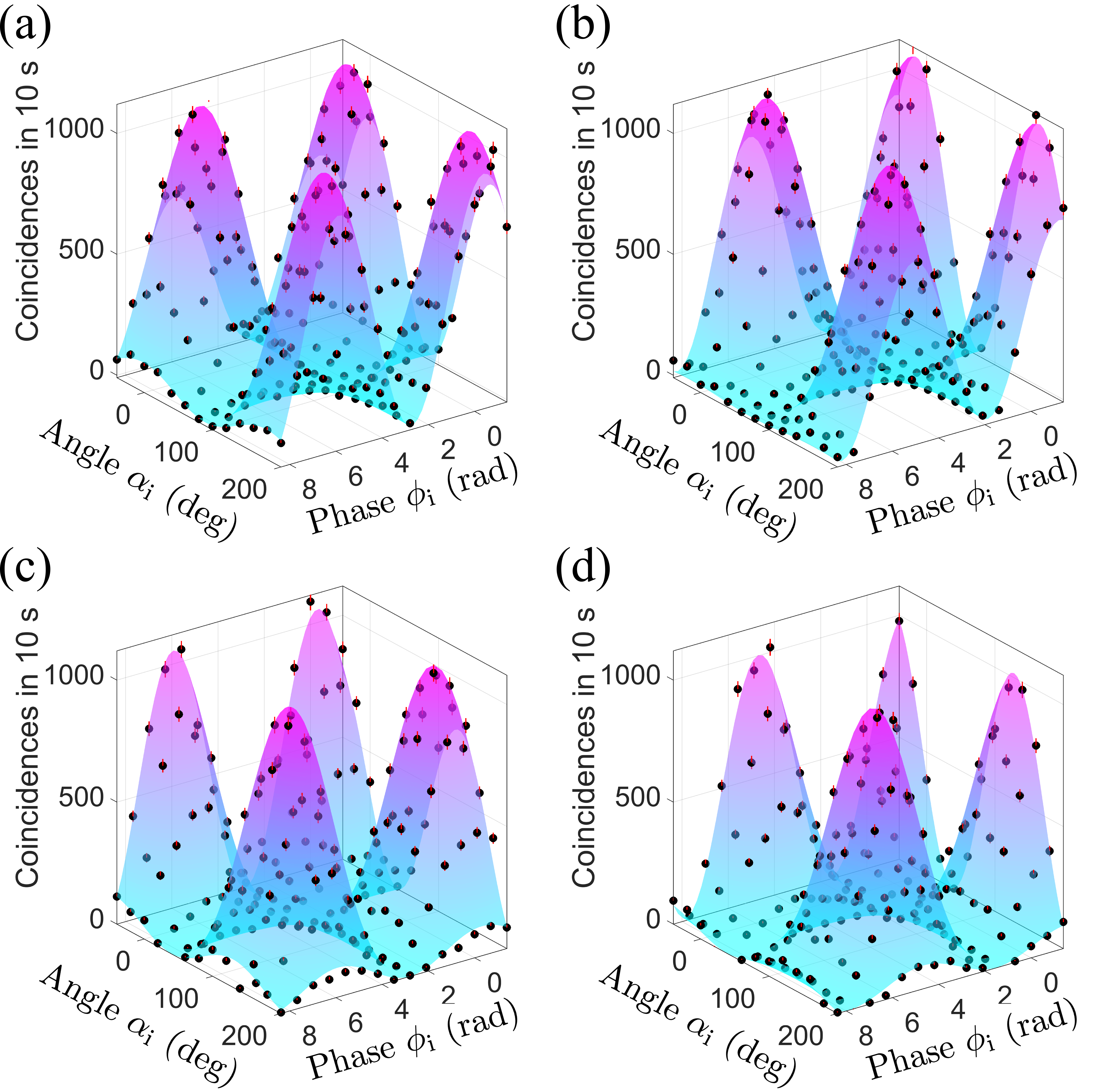}
\caption{\textbf{Coincidence count rates} for four fixed settings at Alice's analyzers: (a) $\{ \alpha_{\rm s} = 0^{\circ}, \phi_{\rm s} = 0 \}$, (b) $\{ \alpha_{\rm s} = 0^{\circ}, \phi_{\rm s}' = \tfrac{\pi}{2} \}$, (c) $\{ \alpha_{\rm s}' = 45^{\circ}, \phi_{\rm s} = 0 \}$, and (d) $\{ \alpha_{\rm s}' = 45^{\circ}, \phi_{\rm s}' = \tfrac{\pi}{2} \}$. For all measurements, Bob's settings $\alpha_{\rm i}$ and $\phi_{\rm i}$ are scanned. Dots represent experimental data. The surface is a least-error-square fit using a 2-dimensional sine function. Note that noise originating only from detector dark counts has been subtracted (on average, 20 dark counts per measurement). \label{surfaces}}
\end{figure}

\begin{center}
\textit{Results}
\end{center}
\figurename~\ref{surfaces} shows the obtained coincidence count rates in the channel pair ITU10$-$33 for four fixed settings at Alice's analyzers ($\{ \alpha_{\rm s} = 0^{\circ}, \phi_{\rm s} = 0 \}$, $\{ \alpha_{\rm s} = 0^{\circ}, \phi_{\rm s}' = \tfrac{\pi}{2} \}$, $\{ \alpha_{\rm s}' = 45^{\circ}, \phi_{\rm s} = 0 \}$, and $\{ \alpha_{\rm s}' = 45^{\circ}, \phi_{\rm s}' = \tfrac{\pi}{2} \}$), and variable settings $\alpha_{\rm i}$ and $\phi_{\rm i}$ at Bob's site.
The results demonstrate the rotation invariance of the correlations as all measurements are essentially similar, except a polarization and/or phase offset equal to the settings $\{ \alpha_{\rm s}', \phi_{\rm s}' \}$ of Alice's analyzers. Experimental data are fitted with a 2-dimensional sine function for which interference fringe visibilities of 98.0\%$\pm$1.5\% are inferred for both polarization and energy-time observables.

\begin{figure}[t]
\includegraphics[width=0.95\columnwidth]{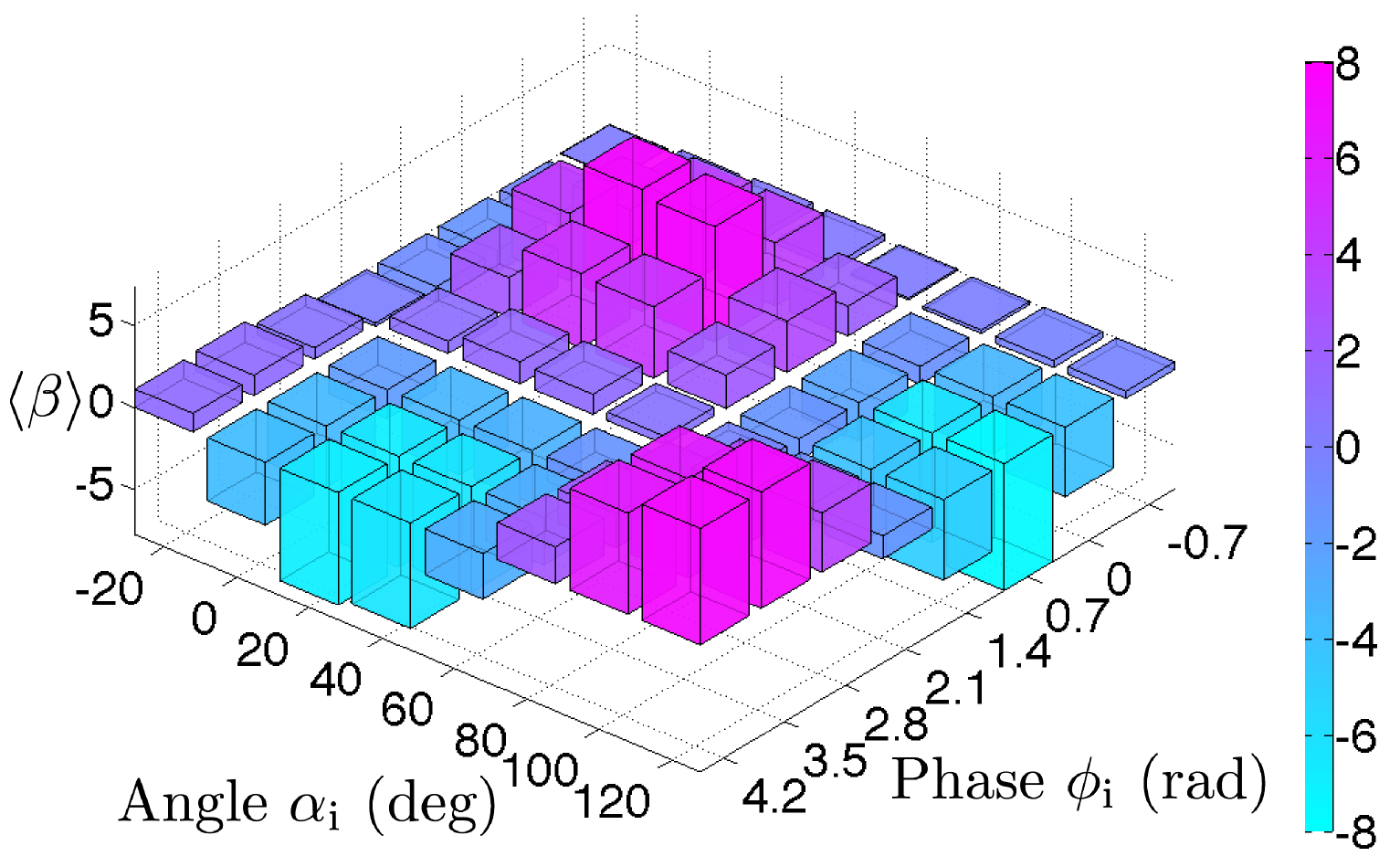}
\caption{\textbf{Violation of the generalized Bell inequality.} $\langle \beta \rangle$ is inferred for different analyzer settings at Bob's site, \textit{i.e.} $\alpha_{\rm i}$ and $\phi_{\rm i}$. The complementary settings are $\alpha_{\rm i}' = \alpha_{\rm i}'  + 45^{\circ}$ and $\phi_{\rm i}' = \phi_{\rm i} + \tfrac{\pi}{2}$. The optimal value for $\langle \beta \rangle$ is obtained at $\{\alpha_{\rm i} = 22.5^{\circ}, \phi_{\rm i} = \tfrac{\pi}{4} \}$. \label{sparopt}}
\end{figure}

We then extract $\langle \beta \rangle$ for different polarization and phase settings $\alpha_{\rm i}$ and $\phi_{\rm i}$ at Bob's site, for which the complementary settings are $\alpha_{\rm i}' = \alpha_{\rm i} + 45^{\circ}$ and $\phi_{\rm i}' = \phi_{\rm i} + \tfrac{\pi}{2}$~\footnote{Note that as only two detectors were available, our measurements represent a 1-outcome filter configuration. We estimated the correlations for the other 15 possible outcomes under the assumption that the overall coincidence rate is constant.}.
The results in \figurename~\ref{sparopt} show several local extrema of $\langle \beta \rangle$. The maximum at $\{\alpha_{\rm i} = 22.5^{\circ}, \phi_{\rm i} = \tfrac{\pi}{4} \}$ amounts to $\langle \beta \rangle = 7.73 \pm 0.12$, thus violating the generalized Bell inequality by 31 standard deviations~\cite{Barbieri_hyper_2006}.
The inferred correlation strengths for all the 16 different combinations of settings are shown in  
\tablename~\ref{expvalues}. Let us stress that for six (ten) combinations of settings, negative (positive) correlations are observed, similarly to standard Bell inequality tests where usually three positive and one negative correlators are found. 
\begin{table}[!h]
\begin{center}
\begin{tabular}{ c | c  c  c  c }  
Settings  & $\{ \alpha_{\rm s},\alpha_{\rm i} \}$ & $\{ \alpha_{\rm s},\alpha_{\rm i}' \}$ & $\{ \alpha_{\rm s}',\alpha_{\rm i} \}$ & $\{ \alpha_{\rm s}',\alpha_{\rm i}' \}$ \\ \hline 
$\{ \phi_{\rm s},\phi_{\rm i} \}$ & 0.51 & -0.33 & -0.46 & -0.41 \\ 
$\{ \phi_{\rm s},\phi_{\rm i}' \}$ & -0.57 & 0.34 & 0.50 & 0.54 \\ 
$\{ \phi_{\rm s}',\phi_{\rm i} \}$ & -0.36 & 0.30 & 0.62 & 0.52 \\ 
$\{ \phi_{\rm s}',\phi_{\rm i}' \}$ & -0.69 & 0.58 & 0.55 & 0.46 \\ 
\end{tabular}
\caption{\label{expvalues} \textbf{Measured correlation strengths} for the 16 measurement setting combinations in the wavelength channel pair ITU$10-23$. The four polarization settings are $\alpha_{\rm s}=0^{\circ}$, $\alpha_{\rm s}'=45^{\circ}$, $\alpha_{\rm i}=22.5^{\circ}$, and $\alpha_{\rm i}'=67.5^{\circ}$. The phase-settings are $\phi_{\rm s} = 0$, $\phi_{\rm s}' = \tfrac{\pi}{2}$, $\phi_{\rm i} = \tfrac{\pi}{4}$, and $\phi_{\rm i}' = \tfrac{3\,\pi}{4}$. Typical uncertainties are of about 0.03.}
\end{center}
\end{table}

Although these high-quality results underline the suitability of our scheme for distributing hyperentanglement, we have to consider that $| \langle \beta \rangle | \approx 8$ can be reached by different mixtures of polarization/energy-time hyperentangled states (and not only the state given in equation~\ref{eq_1}. To this end, we perform an additional check in which we consider our source as a ``black box''. This way, we remove this ambiguity by performing maximum likelihood estimation tomography on the measured data, without making any prior assumptions on the prepared state~\cite{Qtomo}.
By inferring all the density operators associated with the experimentally observed correlations, we extract that the state given in equation~\ref{eq_1} is generated with a net fidelity of $0.86 \leq \mathcal{F} \leq 0.95$ ($0.85 \leq \mathcal{F} \leq 0.94$ without subtraction of detector dark counts). The uncertainty in this measurement is primarily due to limited coincidence statistics. Nevertheless, this is enough to ensure that the state has indeed reached the maximum dimensionality~\cite{Fickler_hyper_2014,Bavaresco_hyper_2018}.

To further demonstrate compliance with DWDM telecom standards, we repeat the measurements in four additional channel pairs. The summary of the experimental results are shown in \tablename~\ref{DWDMresults}.
\begin{table}[!h]
\begin{center}
\begin{tabular}{ c | c | c | c}
Signal channel& Idler channel & Maximum & Standard \\
ITU & ITU & $\langle \beta \rangle$ & deviations \\ \hline
10 & 33 & $7.73 \pm 0.12$ & 31 \\
11 & 32 & $7.25 \pm 0.12$ & 27\\
12 & 31 & $7.63 \pm 0.12$ & 30\\
13 & 30 & $7.61 \pm 0.11$ & 32\\
14 & 29 & $7.78 \pm 0.13$ & 29\\
\end{tabular}
\caption{\textbf{Results in different DWDM channels.} In all channel pairs, a strong violation of the generalized Bell inequality is observed~\cite{Barbieri_hyper_2006}. \label{DWDMresults}}
\end{center}
\end{table}
We observe a clear violation of the generalized Bell inequality in all channel pairs by more than 27 standard deviations, which stands are a clear validation of our approach.

\begin{center}
\textit{Discussion}
\end{center}
Although the above-demonstrated high-quality results are already sufficient to envisage quantum key distribution tasks, additional improvements could be made by further boost the quality of the source. \textit{E.g.}, our current MI active phase-stabilisation system achieves a stability of $\pm 2\pi/40$, which results in a fringe visibility reduction of 1.2\%. The photon pair contribution probabilities for $|H \rangle_{\rm s} |H \rangle_{\rm i}$ and $|V \rangle_{\rm s}|V \rangle_{\rm i}$, fluctuate by $\pm 1\%$, thus leading to a further visibility reduction of 0.2\%. This issue could be overcome by gluing the fibre ends to the PPLN/w chip. All further visibility reduction is explained by non-equilibrated loss in the short and long MI arms, non-ideal polarisation state behaviour of the used HWPs and PBSs, as well as small residual polarisation-dependent loss throughout the setup.

In view of maximising bit rates in quantum networking applications, it is important to note that the use of DWDM filters typically induces an extra 3\,dB of loss per photon. However, as we will show now, the DWDM strategy still achieves higher bit rates. This is because the bottleneck in photonic entanglement distribution is usually the pair generation rate per channel, $R$, which is limited by three constraints: 1.) to avoid degradation due to multi-pair events, the rate is limited to about 5\% of the inverse single-photon coherence time~\cite{Aktas_DWDM_2016} (5\,ps with 5-channel DWDM, 1\,ps without DWDM); 2.) for the same reason, the rate must stay below about 5\% of the inverse timing resolution of the detector; 3.) the detected photon pair rate must stay below the detector saturation.
In our particular setup, count rates are limited by detector saturatation at 20\,kcps. Further, the transmission per photon from source to detector is -13\,dB  (-10\,dB) with DWDM (without DWDM), and we use only two detectors per channel. Thus, the pair generation rate, per channel, at the source is limited to $R_{\rm DWDM} = 8 \cdot 10^6\rm\,s^{-1}$ and $R_{\rm no} = 4 \cdot 10^6\rm\,s^{-1}$, respectively. This leads to detected coincidence rates of $C_{\rm DWDM} = 5 \times 100\rm\,cps$ with DWDM, and $C_{\rm no} = 200\rm\,cps$. The DWDM advantage remains also if using the best-in-class detectors with 90\% efficiency, 15\,ps timing resolution, and saturation at about 150\,Mcps~\cite{Zadeh_detectors_2017}.
With and without DWDM, pair generation rate, per channel, is now limited by the detector timing resolution to $R_{\rm DWDM} = R_{\rm no} = 3.3\cdot 10^9\rm\,s^{-1}$. The expected coincidence rates in our setup would then be $C_{\rm DWDM} = 5 \times 0.83\,\rm Mcps$, and $C_{\rm no} = 3.4\,\rm Mcps$. We note that for long-distance scenarios with significantly higher transmission loss, the DWDM advantage remains at the number of exploited channels divided by the square of the additional single-photon transmission loss, as the detector timing resolution will always represent the ultimate limitation.\\
It may also be interesting to further enhance the bitrates by combining our work with recently demonstrated multi-core fibre distribution techniques~\cite{Ding_Multicore_2017}.

\begin{center}
\textit{Conclusion}
\end{center}
In summary, we have demonstrated high-quality wavelength division multiplexed hyperentanglement generation and analysis. A generalised Bell inequality was violated by more than 27 standard deviations in all channel pairs, and our results were reinforced by maximum likelihood tomography.
The very way that hyperentanglement is generated in our scheme allows it to be straightforwardly adapted to the needs of various experiments across all fields of quantum information science. As examples, a similar source has been used for a fundamental quantum physics experiment~\cite{TBBT_Nature_2018} and for the quantum-enhanced determination of fibre chromatic dispersion~\cite{Vergyris_sagnac_2017}.

We showed further that our scheme can be straightforwardly applied to practical fiber-based quantum key distribution with increased bit rates compared to ordinary schemes~\cite{Aktas_DWDM_2016}. In this perspective, it has already been shown that wavelength division multiplexed quantum key distribution is possible with only a moderate increase in resources~\cite{Kaiser_optimal_2016}.
In view of such an implementation, the presented fully guided-wave scheme further allows ultra-compact and stable design, \textit{e.g.} including a fiber pigtailed PPLN/w module and an integrated PBS.
The performance of such protocols can be further enhanced using quantum memories capable of storing hyperentanglement~\cite{Tiranov_memory_2015}.
The robustness of quantum networks is also increased through hyperentanglement~\cite{Walborn_2008_hyperentanglement}. For example, if one entanglement analyser fails, secure quantum key distribution is still possible by exploiting the other DOFs. \textit{E.g.}, by averaging over all polarisation settings, our net date show that we violate Bell's inequality for energy-time entanglement with $|\langle \beta_1 \rangle | = 2.69 \pm 0.01$. Similarly, when averaging over all energy-time settings, polarisation entanglement is observed with $|\langle \beta_2 \rangle | = 2.72 \pm 0.02$.

We therefore believe that our approach has the potential to become a working horse solution in a large variety of photonics applications where quantum enhancement is sought.


\subsection{Acknowledgements}

The authors acknowledge financial support from the Foundation Simone \& Cino Del Duca, the European Commission for the FP7-ITN PICQUE project (grant agreement No 608062), l'Agence Nationale de la Recherche (ANR) for the e-QUANET, CONNEQT, INQCA, and SPOCQ projects (grants ANR-09-BLAN-0333-01, ANR-EMMA-002-01, ANR-14-CE26-0038, and ANR-14-CE32-0019, respectively), the iXCore Research Foundation, and the French government through its program ``Investments for the Future'' under the Universit\'e C\^ote d'Azur UCA-JEDI project (under the label Quantum$@$UCA) managed by the ANR (grant agreement ANR-15-IDEX-01). This work was also conducted within the framework of the project OPTIMAL granted by the European Union and the R\'egion PACA by means of the Fond Europ\'een de D\'eveloppement R\'egional, FEDER.

\end{document}